\documentclass{elsart}

\usepackage{amssymb}
\usepackage{graphicx}% Include figure files
\usepackage{dcolumn}% Align table columns on decimal point
\usepackage{bm}% bold math
\usepackage{amsmath}
\usepackage{comment}

\begin{document}

\begin{frontmatter}
\title{Criticality in the configuration-mixed interacting boson model:\\
(1) $U(5)$--$\hat{Q}(\chi)\cdot\hat{Q}(\chi)$ mixing}
\author[label1]{V. Hellemans},
\ead{veerle.hellemans@ugent.be}
\author[label2]{P. Van Isacker},
\author[label1]{S. De Baerdemacker}
and
\author[label1,label3]{K. Heyde}
\address[label1]{Department of Subatomic and Radiation Physics,\\
Proeftuinstraat 86, B-9000 Gent, Belgium}
\address[label2]{Grand Acc\'el\'erateur National d'Ions Lourds,
CEA/DSM--CNRS/IN2P3,\\
B.P. 55027, F-14076 Caen Cedex 5, France}
\address[label3]{ISOLDE, CERN, CH-1211 Geneva 23, Switzerland}
\begin{abstract}
The case of $U(5)$--$\hat{Q}(\chi)\cdot\hat{Q}(\chi)$ mixing
in the configuration-mixed Interacting Boson Model
is studied in its mean-field approximation.
Phase diagrams with analytical and numerical solutions
are constructed and discussed.
Indications for first-order and second-order shape phase transitions
can be obtained from binding energies
and from critical exponents, respectively.
\end{abstract}
\begin{keyword}
Interacting boson model \sep
configuration mixing \sep
phase transitions \sep
critical exponents
\PACS 21.60.Fw \sep 21.60.Ev \sep 05.70.Fh \sep 05.70.Jh
\end{keyword}
\end{frontmatter}
\section{Introduction}
The interacting boson model (IBM)
introduced by Arima and Iachello~\cite{iachello87} is an algebraic model
that has its roots in the nuclear shell model.
The approximation of the IBM that only $L=0$ and $L=2$ nucleon pairs
are considered and mapped onto $s$ and $d$ bosons
gives rise to the group structure $U(6)$.
This serves as the dynamical algebra of the model,
{\it i.e.}, the Hamiltonian and other operators
can be expressed in terms of the generators of $U(6)$.
Furthermore, the $U(6)$ group structure
leads to the remarkable property
that the Hamiltonian is analytically solvable
for certain choices of the interaction parameters.
In spite of its microscopic underpinning
in terms of the shell model of the atomic nucleus,
the IBM can also be linked to a macroscopic interpretation of the nucleus
by means of the coherent-state formalism~\cite{ginocchio80,dieperink80,bohr80}.
This formalism allows one to associate an energy surface
in the collective quadrupole shape parameters $\beta$ and $\gamma$
with any IBM Hamiltonian.
Hence, the analytically solvable limits of the Hamiltonian,
the $U(5)$, $O(6)$ and $SU(3)$ limits,
can be linked to a spherical vibrator, a $\gamma$-independent rotor,
and a prolate or oblate deformed rotor, respectively.
The evolution of the IBM Hamiltonian with varying parameters
and the associated energy surface
has been studied extensively~\cite{feng81,lopez96,jolie02,iachello04,leviatan06}.It was shown that the energy surface
undergoes a first-order quantum phase transition
in the passage from $U(5)$ to $SU(3)$
and a second-order quantum phase transition from $U(5)$ to $O(6)$.
The concept of quantum phase transitions
was introduced by Gilmore {\it et al.}~\cite{gilmore78,feng79}
in analogy with the well-known thermodynamic phase transitions.
Quantum phase transitions are not driven
by the control parameter temperature, however,
but rather by the parameters of the Hamiltonian
describing the quantum system.\\
In the IBM in its simplest form
the bosons are restricted to the valence space
but the model can be extended
to a configuration-mixed version (IBM-CM)~\cite{duval81,duval82}
where particle--hole (p--h) excitations
across a closed proton or neutron shell are incorporated.
In certain regions of the nuclear chart
these p--h excitations descend very low in energy
such that they can strongly interact with the regular configuration
or even become the ground state.
Macroscopically this is understood as shape coexistence,
the coexistence of several minima of the energy surface
within a very small energy interval.
This macroscopic information can be extracted from the IBM-CM
by calculating the expectation value in a coherent state
appropriate for configuration mixing~\cite{frank04}.
The resulting energy surface exhibits a single minimum
or several coexisting minima depending on the IBM-CM parameters.
Recently, the energy surface for $U(5)$--$O(6)$ mixing
has been studied~\cite{frank06}
and it was shown that the IBM-CM in this case
gives rise to an extended phase with shape coexistence.\\
The aim of this paper is the study
of the more general case of $U(5)$--$\hat{Q}(\chi)\hat{Q}(\chi)$ mixing,
where $\hat{Q}(\chi)$ (see sect.~\ref{sect:ibm}) is the quadrupole operator
which drives the system to deformation
and $U(5)$ is the spherical-vibrator limit of the IBM.
\section{The energy surface
for $U(5)$--$\hat{Q}(\chi)\hat{Q}(\chi)$ mixing}
\label{sect:ibm}
The most compact form of the IBM Hamiltonian,
which captures the essential physics of the model,
is obtained within the consistent-$Q$ formalism~\cite{warner83}
\begin{equation}
\hat{H}_{\rm cqf}=
\epsilon\hat{n}_d
-|\kappa|\hat{Q}(\chi)\cdot\hat{Q}(\chi)~,
\label{eq:Hibm}
\end{equation}
where $\hat{n}_d$ is the $d$-boson number operator and $\hat{Q}(\chi)=(s^\dag\tilde{d}+d^\dag s)^{(2)}+
\chi(d^\dag\tilde{d})^{(2)}$ the quadrupole operator.
For specific choices of the parameters
$\epsilon$, $\kappa$ and $\chi$,
the three symmetry limits of the IBM are obtained:
the $U(5)$ limit for $\kappa$=0,
the $O(6)$ limit for $\epsilon=0$ and $\chi=0$
and the $SU(3)$ limit for $\epsilon=0$ and $\chi=\pm\sqrt{7}/2$.
By calculating the expectation value of the Hamiltonian~(\ref{eq:Hibm})
in a normalised projective coherent state~\cite{ginocchio80,dieperink80,bohr80}
\begin{equation}
|N,\beta,\gamma\rangle=
\frac{1}{\sqrt{N!}(1+\beta^2)^{N/2}}
\left(s^\dag+\beta\left[\cos\gamma~d_0^\dag+ \frac{1}{\sqrt{2}}\sin\gamma(d^\dag_2+d^\dag_{-2})\right]\right)^N|0\rangle~,
\label{eq:coherentstate}
\end{equation}
the associated energy surface is obtained
\begin{eqnarray}
E^N(\epsilon,|\kappa|,\chi,\beta,\gamma)&\equiv&
\langle N,\beta, \gamma|\hat{H}_{\rm cqf}|
N,\beta,\gamma\rangle
\nonumber\\&=&
\epsilon N\frac{\beta^2}{1+\beta^2}-|\kappa|\left[\frac{N\left[5+(1+\chi^2)\beta^2\right]}{1+\beta^2}\right.
\nonumber\\&+&
\left.\frac{N(N-1)}{(1+\beta^2)^2}
\left(\frac{2}{7}\chi^2\beta^4-
4\sqrt{\frac{2}{7}}\chi\beta^3\cos(3\gamma)+
4\beta^2 \right)\right]~,
\end{eqnarray}
where $N$ denotes the number of valence bosons
and ($\beta,\gamma$) are collective variables.
If the values of the parameters
for the three different IBM limits are inserted,
it is found that the $U(5)$ limit
can be associated with an energy surface with a spherical minimum,
the $O(6)$ limit with one
with a deformed but $\gamma$-independent minimum,
and the $SU(3)$ limit with an energy surface
which has either a prolate (for $\chi=-\sqrt{7}/2$)
or an oblate (for $\chi=\sqrt{7}/2$) deformed minimum.\\
An extended version of the IBM with configuration mixing (IBM-CM)
allows the simultaneous treatment and mixing
of several boson configurations
which correspond to different particle--hole (p--h)
shell-model excitations~\cite{duval81,duval82}.
In particular, configurations with $N$, $N+2$, $N+4$, \dots bosons
are associated with 0p--0h, 2p--2h, 4p--4h, \dots excitations, respectively.
In case of mixing between a `regular' 0p--0h and a `intruder' 2p--2h configuration,
the Hamiltonian can be written as
\begin{equation}
\hat{H}=
\hat{P}^{\dag}_{N}\hat{H}^N_{\rm cqf}\hat{P}_{N}+
\hat{P}^{\dag}_{N+2}\left(\hat{H}^{N+2}_{\rm cqf}+\Delta\right)\hat{P}_{N+2}+
\hat{V}_{\rm mix}~,
\end{equation}
where $\hat{P}_{N}$ and $\hat{P}_{N+2}$ are operators
projecting onto the $N$-boson and $(N+2)$-boson spaces, respectively.
Although the Hamiltonians are formally equivalent,
the different superscripts
in $\hat{H}^{N}_{\rm cqf}$ and $\hat{H}^{N+2}_{\rm cqf}$
indicate that the parametrisation can be configuration dependent.
The parameter $\Delta$ is the energy needed
to excite two particles across a shell gap,
corrected for the pairing interaction and a monopole effect~\cite{heyde87}.
Finally, $\hat{V}_{\rm mix}\equiv
w_0 (s^\dag s^\dag+ss)+
w_2 (d^\dag\cdot d^\dag+\tilde{d}\cdot\tilde{d})$
denotes the interaction between the two configurations.
This form of the Hamiltonian can be easily extended
to incorporate configuration mixing
with higher-order particle--hole excitations.\\
The geometric interpretation of the IBM-CM is obtained
by introducing a matrix coherent-state method~\cite{frank04}.
In case of mixing between a 0p--0h and a 2p--2h configuration
the energy surface is given by the lowest eigenvalue of the matrix
\begin{equation}
\left[
\begin{array}{cc}
E^N(\epsilon_1,|\kappa_1|,\chi_1,\beta,\gamma)&\omega(\beta)\\
\omega(\beta)&E^{N+2}(\epsilon_2,|\kappa_2|,\chi_2,\beta,\gamma)+\Delta
\end{array}\right]~,
\end{equation}
with
\begin{equation}
\omega(\beta)\equiv
\langle N,\beta,\gamma|
\hat{V}_{\rm mix}|
N+2,\beta,\gamma\rangle=
\sqrt{(N+2)(N+1)}\;\frac{w_0+w_2\beta^2}{1+\beta^2}~,
\end{equation}
and $E^N(\epsilon_1,|\kappa_1|,\chi_1,\beta,\gamma)$
and $E^{N+2}(\epsilon_2,|\kappa_2|,\chi_2,\beta,\gamma)$
the expectation values
of $\hat{H}^{N}_{\rm cqf}$ and $\hat{H}^{N+2}_{\rm cqf}$
in the appropriate projective coherent state.
For simplicity's sake, $w_0$ and $w_2$
are taken equal ($w_0=w_2\equiv w$)
such that $\omega$ becomes $\beta$ independent.\\
In the language of catastrophe theory~\cite{gilmore81},
which can be used to study qualitative changes in the energy surface, 
$(N,\epsilon_1,|\kappa_1|,\chi_1,\epsilon_2,|\kappa_2|,\chi_2,\Delta,\omega)$
are called control parameters.
Because a general study of the energy surface
as a function of all nine control parameters
is totally out of (current) computational reach,
we focus on mixing between the dynamical symmetries of the IBM
which can be considered as the benchmarks of the model.
In the present paper we concentrate on the case of mixing
between the vibrational $U(5)$ limit
and the deformation driving quadrupole term $\hat{Q}(\chi)\cdot\hat{Q}(\chi)$ which incorporates both the $O(6)$ limit
and the $SU(3)$ limit for $\chi=0$ and $\chi=\pm \sqrt{7}/2$, respectively.
In a forthcoming paper the case of mixing
between deformed configurations will be treated.\\
The energy surface resulting from the matrix coherent-state method
in the case of $U(5)$--$\hat{Q}(\chi)\cdot\hat{Q}(\chi)$ mixing
is given by
\begin{eqnarray}
E_-&=&
\frac{|\kappa|}{2(1+\beta^2)^2}
\Bigg(\left[\epsilon' N-(N+2)(1+\chi^2)-\frac{2}{7}(N+2)(N+1)\chi^2+\Delta'\right]\beta^4
\nonumber\\
&&+\left[\epsilon' N-(N+2)(6+\chi^2)-4(N+2)(N+1)+2\Delta'\right]\beta^2
\nonumber\\
&&+\frac{4}{7}(N+2)(N+1)\sqrt{14}\chi\beta^3\cos(3\gamma)-5(N+2)+\Delta'
\nonumber\\
&&-\bigg[\bigg(
\left[\epsilon' N+(N+2)(1+\chi^2)+\frac{2}{7}(N+2)(N+1)\chi^2-\Delta'\right]\beta^4
\nonumber\\
&&\quad+\left[\epsilon' N+(N+2)(6+\chi^2)+4(N+2)(N+1)-2\Delta'\right]\beta^2
\nonumber\\
&&\quad-\frac{4}{7}(N+2)(N+1)\sqrt{14}\chi\beta^3\cos(3\gamma)
+5(N+2)-\Delta'\bigg)^2
\nonumber\\
&&\quad+\omega'^2(1+\beta^2)^4\bigg]^{\frac{1}{2}}\Bigg)~,
\label{eq:energysurface}
\end{eqnarray}
where $\Delta'=\Delta/|\kappa|$, $\epsilon'=\epsilon/|\kappa|$
and $\omega'=2\omega/|\kappa|$.
We will omit the scaling factor $|\kappa|/2$ from now on
as the structural properties only depend on $\Delta'$, $\epsilon'$ and $\omega'$.
\section{Criticality conditions and Maxwell points}
\subsection{Introduction}
In the following we rely on the ideas of catastrophe theory
as discussed extensively by Gilmore~\cite{gilmore81}.
In the family of energy surfaces
$E_-(\beta,\gamma; \epsilon',\chi,\Delta',\omega',N)$ under study,
$(\epsilon',\chi,\Delta',\omega',N)$ are referred to as the control parameters
while $(\beta,\gamma)$ are the collective variables.
In general, for an arbitrary set of control parameters $(\epsilon_0',\chi_0,\Delta_0',\omega_0',N_0)$,
the energy surface in $(\beta,\gamma)$
exhibits isolated critical (or equilibrium) points.
Isolated critical points are characterised
by a vanishing gradient of the energy surface ($\nabla E_-=0$)
and a non-zero determinant of the stability matrix
($\det(\mathcal{S})\neq$ 0).
The latter matrix is defined as
\begin{equation}
\mathcal{S}=
\left[\begin{array}{ccc}
{\displaystyle\frac{\partial^2 E_-}{\partial\beta^2}}&~&
{\displaystyle\frac{\partial^2 E_-}{\partial\beta\partial\gamma}}\\
{\displaystyle\frac{\partial^2 E_-}{\partial\gamma\partial\beta}}&~&
{\displaystyle\frac{\partial^2 E_-}{\partial\gamma^2}}
\end{array}\right]~,
\end{equation}
and its eigenvalues determine the stability properties of the energy surface
in isolated critical points.
If all the eigenvalues of $\mathcal{S}$ are positive,
the isolated critical point is a minimum;
negative eigenvalues of $\mathcal{S}$ indicate a maximum
whereas positive and negative eigenvalues characterise a saddle point.
Since the isolated critical points
are the extrema or saddle points of the energy surface
$E_-(\beta,\gamma; \epsilon_0',\chi_0,\Delta_0',\omega_0',N_0)$,
they determine its global behaviour.
%
%\begin{comment}
\begin{figure}%[!htb]
\begin{center}
\includegraphics{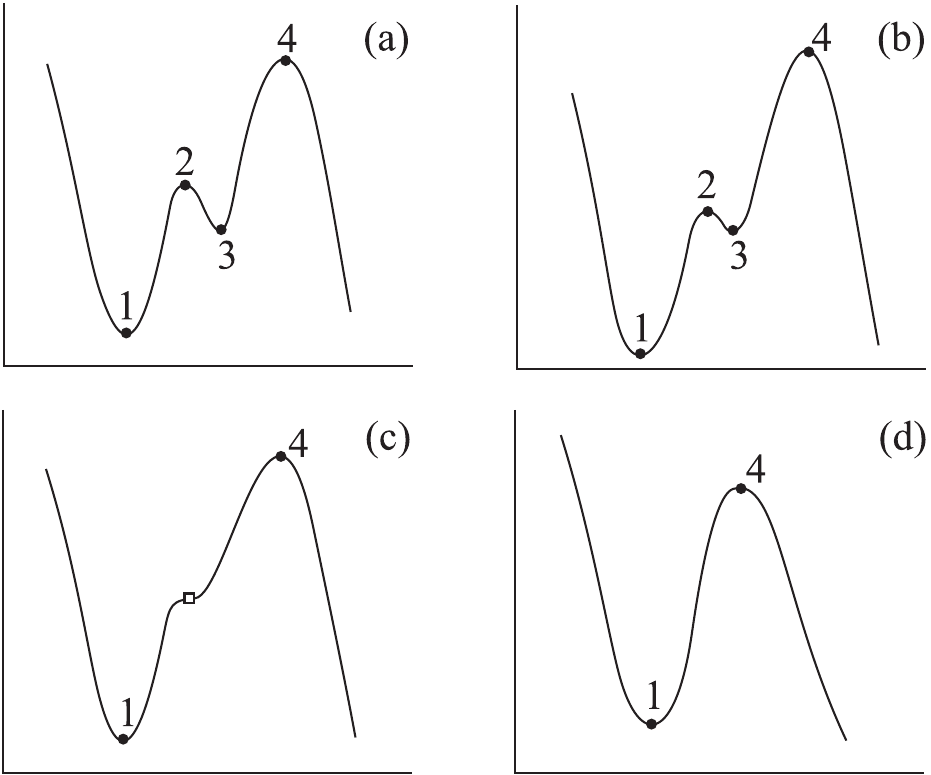}
\end{center}
\caption{Illustration of the fact that degenerate critical points
organise the qualitative behaviour of a family of functions.
Panels (a), (b), (c) and (d) show an arbitrary function
for different sets of the control parameters.
In panels (a), (b) and (d) the function exhibits only isolated critical points
(indicated with a dot)
while in panel (c) it has a degenerate critical point
(indicated with a square).
The number of isolated critical points or extrema changes
when the control parameters pass through (c)
corresponding to a degenerate critical point.}
\label{fig:degenerate}
\end{figure}
%\end{comment}
%
For specific values of the control parameters
the energy surface exhibits points
where the determinant of the stability matrix vanishes (det($\mathcal{S}$)=0).
These are the {\em degenerate} critical points
and they are of great importance.
Whereas isolated critical points
organise the qualitative behaviour of a single energy surface,
degenerate critical points organise the qualitative behaviour
of the entire family of energy surfaces
$E_-(\beta,\gamma; \epsilon',\chi,\Delta',\omega',N)$.
If the control parameters are varied and pass through values
where the energy surface exhibits degenerate critical point(s),
the topology of the surface changes.
This can be understood intuitively
by realising that the topology of an energy surface
is determined by its isolated critical points.
Consequently, if two or more isolated critical points
merge into a single degenerate one,
the topology of the energy surface changes.
This is illustrated in fig.~\ref{fig:degenerate}
where the evolution of an arbitrary function
with varying control parameters is shown.
In panel (a) the function exhibits 4 isolated critical points (indicated with a dot).
If the control parameters are changed,
two extrema (2 and 3) move towards each other (panel (b))
until they merge into a degenerate critical point
(indicated with a square) in panel (c).
In panel (d) the degenerate critical point has disappeared
and only two extrema remain.
It is clear that (c) with its degenerate critical point
separates the region where the function exhibits 4 extrema
from the region where it has only 2.\\
Summarising, the degenerate critical points
mark out the different regions in the control parameter space
where the qualitative properties of the energy surface remain unchanged.
Hence, they determine specific lines in the phase diagram.
If two such lines in the phase diagram intersect,
the degeneracy of the crossing point
is higher than the degeneracy of the critical points
determining the lines in the phase diagram.
This crossing is called a triple point.\\
In regions of the phase diagram
where the energy surface has several minima,
it is of interest to know which of these is the global minimum
and where it jumps from one minimum to another
({\it i.e.}, where two degenerate global minima occur).
The locus of points in control parameter space
where this jump of the global minimum occurs
is called the set of Maxwell points.
In the case of $U(5)$--$\hat{Q}(\chi)\cdot\hat{Q}(\chi)$ mixing,
the Maxwell points are the solutions of
\begin{equation}
\left.\frac{\partial E_-}{\partial\beta}\right|_{\beta=\beta_0}=0~,
\qquad
E_-(\beta=\beta_0)-E_-(\beta=0)=0~.
\label{eq:maxconditions}
\end{equation}
\subsection{Analytical solution in $(\beta,\gamma)=(0,n\pi/3)$}
\label{sec:analytical}
In general, the criticality conditions
\begin{equation}
\frac{\partial E_-}{\partial \beta}=0~,
\qquad
\frac{\partial E_-}{\partial \gamma}=0~,
\qquad
\det(\mathcal{S})=0~,
\label{eq:ccondition}
\end{equation}
have to be solved numerically.
However, an analytical solution can be found. Since mixing between spherical $U(5)$
and deformed $\hat{Q}(\chi)\cdot\hat{Q}(\chi)$ is considered,
we expect important changes in the energy surface to occur at $\beta=0$
which is a minimum in the spherical case
and a maximum for the deformed case.
Expanding the energy surface $E_-$ of eq.~(\ref{eq:energysurface})
around $(\beta,\gamma)=(0,n\pi/3)$ ($n$ integer),
we find
\begin{equation}
E_-=
t_{00}+
\frac{1}{2!}t_{20}\beta^2+
\frac{1}{3!}t_{30}\beta^3+
\frac{1}{4!}t_{40}\beta^4+
\frac{1}{5!}t_{50}\beta^5+
\frac{1}{12}t_{32}\beta^3\gamma^2+
\frac{1}{6!}t_{60}\beta^6+\cdots
\label{eq:taylor1}~,
\end{equation}
with
\begin{eqnarray}
t_{00}&=&
-\zeta-\sqrt{\zeta^2+\omega'^2}~,
\nonumber\\
t_{20}&=&
\frac{2}{\sqrt{\zeta^2+\omega'^2}}
\Big[ \epsilon' N\left(\sqrt{\zeta^2+\omega'^2}-\zeta\right)
\nonumber\\&&
-(N+2)(\chi^2+4N)\left(\sqrt{\zeta^2+\omega'^2}+\zeta\right)\Big]~,
\nonumber\\
t_{30}&=& \frac{24}{7}(N+2)(N+1)\sqrt{14}\chi\cos(n\pi)
\frac{\zeta+\sqrt{\zeta^2+\omega'^2}}{\sqrt{\zeta^2+\omega'^2}}~,
\nonumber\\
t_{40}&=&
-24\frac{\zeta+\sqrt{\zeta^2+\omega'^2}}{\sqrt{\zeta^2+\omega'^2}}
\left(\epsilon' N +(N+2)\left[\frac{1}{7}(2N-5)\chi^2-4(2N+1)\right]\right)
\nonumber\\
&&-\frac{12\omega'^2}{\left(\zeta^2+\omega'^2\right)^{\frac{3}{2}}}
\left[\epsilon' N + (N+2)(\chi^2+4N)\right]^2
+\frac{48\zeta\epsilon' N}{\sqrt{\zeta^2+\omega'^2}}~,
\nonumber\\
t_{50}&=&
\frac{480}{7}(N+2)(N+1)\sqrt{14}\chi\cos(n\pi)
\nonumber\\&&
\times\left[-\frac{2(\zeta+\sqrt{\zeta^2+\omega'^2})}
{\sqrt{\zeta^2+\omega'^2}}
+\omega'^2\frac{\epsilon' N+(N+2)(\chi^2+4N)}
{(\zeta^2+\omega'^2)^{\frac{3}{2}}}\right]~,
\nonumber\\
t_{32}&=&-\frac{216}{7}(N+2)(N+1)\sqrt{14}\chi\cos(n\pi)
\frac{\zeta+\sqrt{\zeta^2+\omega'^2}}{\sqrt{\zeta^2+\omega'^2}}~,
\end{eqnarray}
where the
notation $\zeta=-\Delta'+5(N+2)$ is used.\\
The criticality conditions~(\ref{eq:ccondition})
are automatically fulfilled in the point $(\beta,\gamma)=(0,n\pi/3)$
as the linear terms in $\beta$ and $\gamma$
as well as the quadratic term $\gamma^2$ are zero.
Consequently, the lines determining the phase diagram
in control parameter space
for $(\beta,\gamma)=(0,n\pi/3)$
are found by requiring a vanishing $\beta^2$ term in the Taylor expansion.
Hence, if $t_{20}=0$, the stability matrix $\mathcal{S}$ vanishes identically
and we obtain the locus of four-fold degenerate critical points
(two-fold in $\beta$ and two-fold in $\gamma$),
\begin{equation}
\epsilon'_{\rm c}=
-\frac{(N+2)(4N+\chi^2)}{N}
\frac{\zeta+\sqrt{\zeta^2+\omega'^2_{\rm c}}}
{\zeta-\sqrt{\zeta^2+\omega'^2_{\rm c}}}~.
\label{eq:2fold}
\end{equation}
In case of $U(5)$--$O(6)$ mixing one has $\chi=0$
and the energy surface~(\ref{eq:energysurface})
will exhibit a $\beta^4$ behaviour around $(\beta,\gamma)=(0,n\pi/3)$,
if the control variables are chosen according to (\ref{eq:2fold}).
In all other cases the behaviour of the energy surface
at the critical points~(\ref{eq:2fold}) is of dominant $\beta^3$ character.
Note that the global behaviour of the analytical critical line
remains essentially unchanged when the intruder configuration
changes from a $\gamma$-independent rotor
to a prolate/oblate rotor as $\chi$ is part of a positive scaling factor.
If $\zeta>0$ or $\Delta'< 5(N+2)$,
the curve converges asymptotically
to the lines $\omega'=0$ and $\epsilon'=(N+2)(4N+\chi^2)/N$.
If $\Delta'=5(N+2)$, $\epsilon_c'$ takes on
the constant value $(N+2)(4N+\chi^2)/N$.
If $\zeta<0$ or $\Delta'>5(N+2)$,
only the asymptote $\epsilon'=(N+2)(4N+\chi^2)/N$ remains.
As long as $\epsilon'<\epsilon'_{\rm c}$,
a deformed minimum is found
whereas the energy surface
exhibits a spherical minimum for $\epsilon'>\epsilon'_{\rm c}$.
If the excitation energy of the intruder state goes to infinity,
$\Delta'\rightarrow\infty$,
eq.~(\ref{eq:2fold}) reduces to $\epsilon'_{\rm c}=0$.\\
In order to find higher-order degenerate critical points,
higher-order terms are required to vanish.
The coefficient $t_{30}$ vanishes for $\chi=0$,
or  for $\omega'=0$ and $\Delta'>5(N+2)$,
or for $\Delta'=\infty$.
For $\chi=0$ the coefficients $t_{32}$, $t_{50}$ and all higher-order terms
with $\gamma$ dependence disappear.
This can also be seen from the energy surface~(\ref{eq:energysurface})
which becomes $\gamma$ independent if $\chi=0$.
If we additionally impose that $t_{40}=0$,
we find the triple point
\begin{multline}
(\epsilon'_{\rm t},\omega'_{\rm t})=\\
\left(\frac{4N^2(N+2)+\zeta(N+1)}{N^2},
\pm\frac{4N\sqrt{(N+2)[4N^2(N+2)+\zeta(N+1)]}}{N+1}\right)
\label{eq:triplepoint}~.
\end{multline}
In the ($\epsilon',\omega'$) plane of the control parameter space,
eq.~(\ref{eq:triplepoint}) gives the triple point
where the analytical solution~(\ref{eq:2fold})
for $(\beta,\gamma)=(0,n\pi/3)$
and the numerical solution of eq.~(\ref{eq:ccondition})
for $(\beta,\gamma)\neq(0,n\pi/3)$ intersect.
From the Taylor expansion it is seen
that the energy surface exhibits a $\beta^6$ behaviour
in the vicinity of the triple point.
\subsection{Numerical solutions}
In the general case $(\beta,\gamma)\neq (0,n\pi/3)$,
the critical points (see eq.~(\ref{eq:ccondition}))
and the Maxwell points (see eq.~(\ref{eq:maxconditions}))
must be calculated numerically
for the energy surface $E_-$~ in eq.~(\ref{eq:energysurface}).
To simplify the numerical treatment,
$\gamma$ can be ``frozen" to a certain $n\pi/3$.
This follows from the fact
that the condition $\partial E_-/\partial \gamma=0$
implies necessarily that $\chi=0$, $\gamma=n\pi/3$ or $\beta=0$.
The case $\beta=0$ has already been treated
in the analytical solution (sect.~\ref{sec:analytical})
and all $\gamma$ dependence disappears if $\chi=0$.
If we choose $\gamma=n\pi/3$,
$\partial^2 E_-/\partial \gamma^2$ is positive definite
as long as $\gamma$ and the sign of $\chi$ and $\beta$
are chosen consistently.\footnote{
If $\gamma$=0, $\chi$ must be negative and $\beta\geq 0$.
If $\gamma=\pi/3$, a positive sign for $\chi$ and $\beta$ must be chosen.}
Hence, $\gamma$ can be fixed to $n\pi/3$ without loss of generality
and the criticality conditions reduce to
\begin{equation}
\frac{\partial E_-}{\partial\beta}=
\frac{\partial^2 E_-}{\partial\beta^2}=0~.
\end{equation}
In the following sections, we choose $\gamma=0$.
\section{Phase diagrams for $U(5)$--$Q(\chi)\cdot Q(\chi)$ mixing}
%
%\begin{comment}
\begin{figure}[!htb]
\begin{center}
\includegraphics[width=\textwidth]{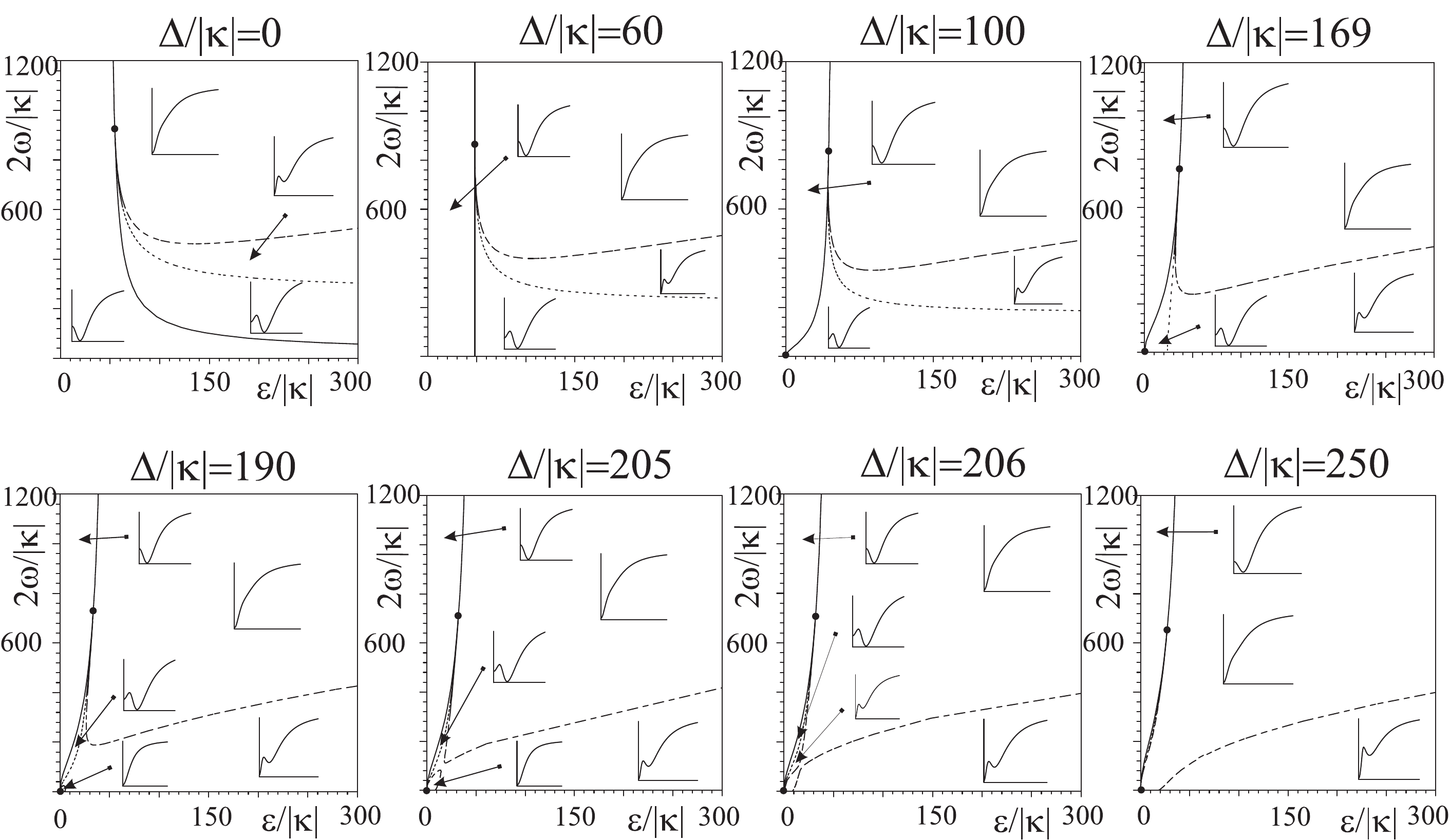}
\end{center}
\caption{Phase diagrams in the case of $U(5)$--$O(6)$ mixing
for several values of $\Delta'$ and for $N=10$.
The locus of analytical critical points is shown as a full line,
that of numerical critical points as a dashed line
and that of Maxwell critical points as a dotted line.
The dots represent triple points. 
The inset figures illustrate the generic shape of the potential
as a function of $\beta$ in each of the zones of the parameter space.}
\label{fig:phasediagram1}
\end{figure}
%\end{comment}
%
In sect.~\ref{sec:analytical} we have shown that
$\chi$ is part of a positive scaling factor
in the analytical solution of the criticality conditions.
The criticality conditions cannot be solved in general
if $\chi$ is considered as a symbolic parameter
but they can if a specific value of $\chi$ is taken.
In this section we discuss the two benchmark cases,
namely $U(5)$--$O(6)$ and $U(5)$--$SU(3)$ mixing.
\subsection{$U(5)$--$O(6)$ mixing}
%
%\begin{comment}
\begin{figure}
\includegraphics[width=\textwidth]{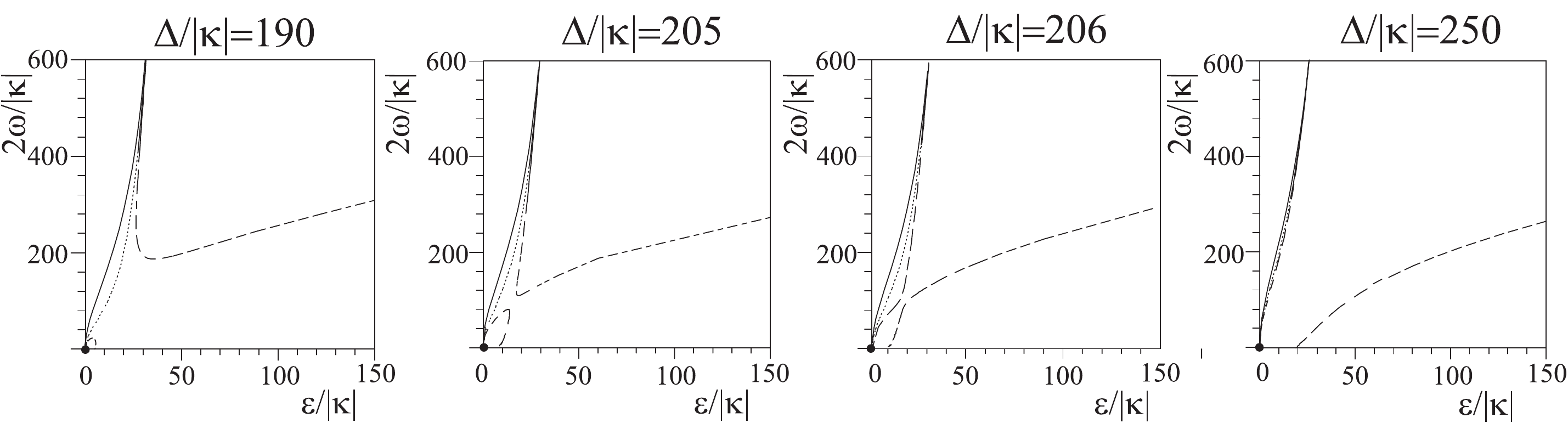}
\caption{Zoom on the part near the origin
of the phase diagrams shown in fig.~\ref{fig:phasediagram1}.
The locus of analytical critical points is shown as a full line,
that of numerical critical points as a dashed line
and that of Maxwell critical points as a dotted line.}
\label{fig:phasediagram1-zoom}
\end{figure}
%\end{comment}
%
The case of $U(5)$--$O(6)$ mixing is obtained
by choosing $\chi=0$ in eq.~(\ref{eq:energysurface}).
In figs.~\ref{fig:phasediagram1} and~\ref{fig:phasediagram1-zoom}
different phase diagrams are shown for several values of $\Delta'$.
The number of bosons is $N=10$.
For $\Delta'=0$ we note a deformed region
to the left of the analytical solution (full line),
a spherical region to the right and above the numerical solution (dashed line)
and a region of shape coexistence in between the two lines.
These three regions meet in the triple point, indicated by a dot.
As $\Delta'$ increases, the slope of the analytical solution switches sign
and a second triple point in the origin is created.
The first triple point moves down the analytical curve
until it merges with the triple point
in the origin ($\epsilon'_{\rm t},\omega'_{\rm t}$)=(0,0)
for $\Delta'=[4N^2(N+2)/(N+1)]+5(N+2)$ (see eq.~(\ref{eq:triplepoint}))
and eventually disappears.
When the slope of the Maxwell curve (dotted line) switches sign,
a small spherical region for low $\epsilon'$ values
starts growing around the origin.
As $\Delta'$ increases further,
the two spherical regions approach each other.
When these two regions coalesce,
the region of shape coexistence is split in two.
The small coexistence region disappears when the two triple points merge,
while the large coexistence region with a spherical global minimum
shifts towards higher $\epsilon'$ for increasing $\Delta'$.
The evolution of the phase diagrams is similar for all $N$,
although the value of $\Delta'$ where changes occur
varies slightly with $N$.
\subsection{$U(5)$--$SU(3)$ mixing}
%
%\begin{comment}
\begin{figure}[!htb]
\begin{center}
\includegraphics [width=\textwidth]{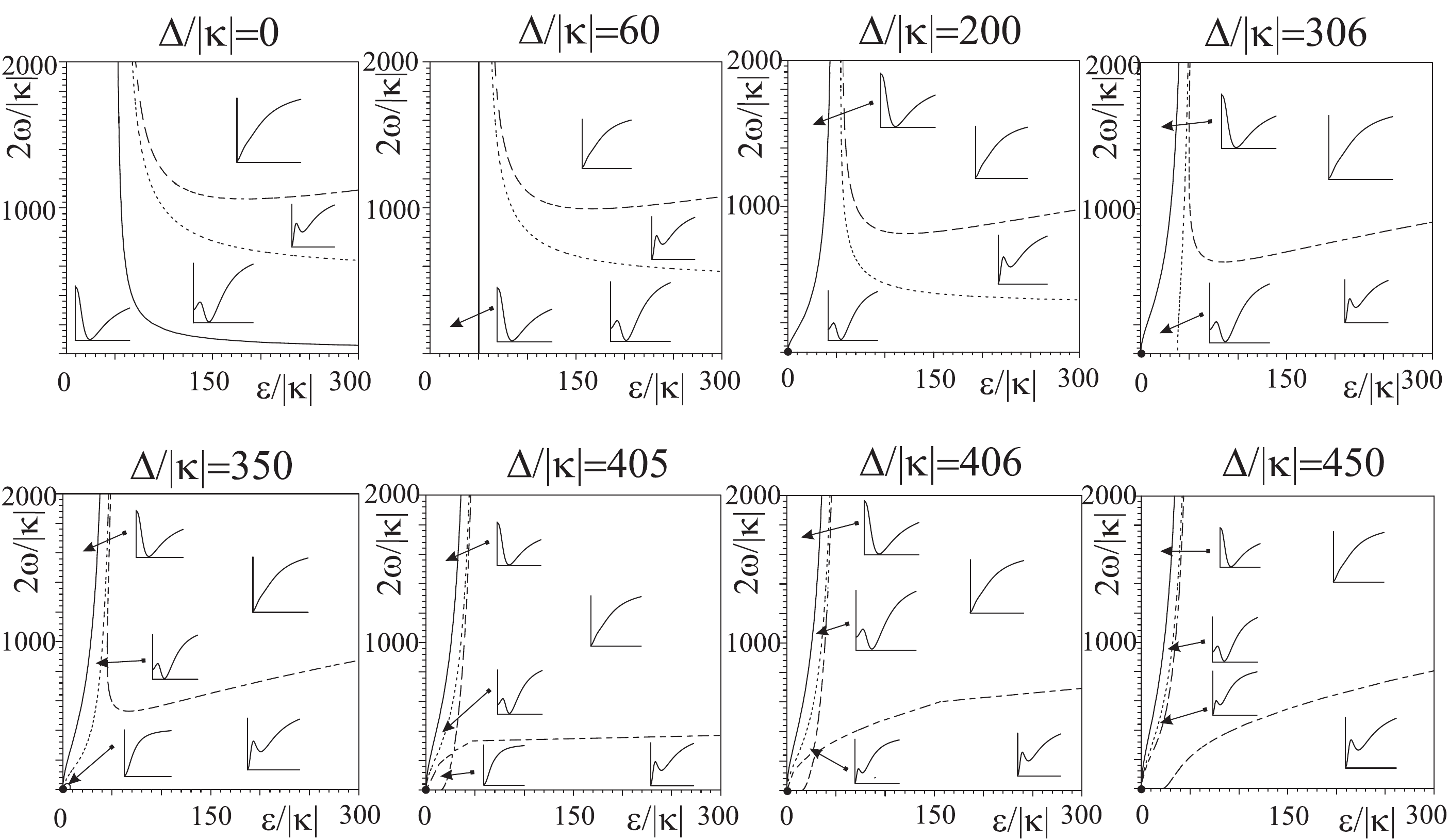}
\end{center}
\caption{Phase diagrams in the case of $U(5)$--$SU(3)$ mixing
for several values of $\Delta'$ and for $N=10$.
The locus of analytical critical points is shown as a full line,
that of numerical critical points as a dashed line
and that of Maxwell critical points as a dotted line.
The inset figures illustrate the generic shape of the potential
as a function of $\beta$ in each of the zones of the parameter space.}
\label{fig:phasediagram2}
\end{figure}
%\end{comment}
%
The phase diagrams obtained
in the case of $U(5)$-$SU(3)$ mixing ($\chi=-\sqrt{7}/2$),
shown in fig.~\ref{fig:phasediagram2},
are very similar to those obtained for $U(5)$--$O(6)$ mixing.
It is important to note the following major differences, however.
As the non-trivial triple point ($\omega'_{\rm t}\neq 0$)
has been proven to occur only for $\chi=0$,
the critical (dashed and full) lines separating the different regions
never meet in a triple point.
This has consequences for the occurrence of shape coexistence
at realistic values of the parameters
when the intruder states lie very high in excitation energy.
In principle, the small region of shape coexistence for low $\epsilon'$ values will only disappear if $\Delta'$ goes to infinity.
Hence, even for very high excitation energies of the intruder states,
there will always be a region of shape coexistence, however small,
for realistic values of $\epsilon'$ and $\omega'$.
This is an essential difference with the case of $U(5)$--$O(6)$ mixing.
\section{Phase transitions}
Similar to the Ehrenfest classification for thermodynamic phase transitions,
a classification for shape or quantum phase transitions
has been proposed \cite{gilmore79}.
The criterion involves the energy of the global minimum $E_{\rm min}$
as a function of a control parameter.
A shape phase transition is called of zeroth order
if $E_{\rm min}$ changes discontinuously at the critical point.
If the first derivative of $E_{\rm min}$ with respect to the order parameter
or its second derivative is discontinuous at the critical point,
the shape phase transition is of first or second order, respectively.
The first-order phase transitions
are characterised by mixed phase regimes,
{\it i.e.} regimes where different phases coexist during the transition.
Typical for second-order phase transitions
is the transition from an ordered to a disordered phase or {\it vice versa}.
\subsection{First-order phase transitions}
It has been shown 
that the energy surface associated with the IBM (without configuration mixing) undergoes a first-order shape phase transition
in the passage from $U(5)$ to $SU(3)$ \cite{dieperink80,dieperink80bis}.
This first-order shape phase transition occurs when passing
through the small region of shape coexistence along the transition path.
However, this type of shape coexistence
is different from the one occurring in the phase diagrams shown here,
as the latter is the result of mixing
between regular and intruder configurations.
Nevertheless, although the underlying physics differs,
the similarities in the associated topology suggest
that first-order phase transitions should also occur here
when passing through the zone of shape coexistence.\\
%
%\begin{comment}
\begin{figure}%[!t]
\includegraphics [width=\textwidth]{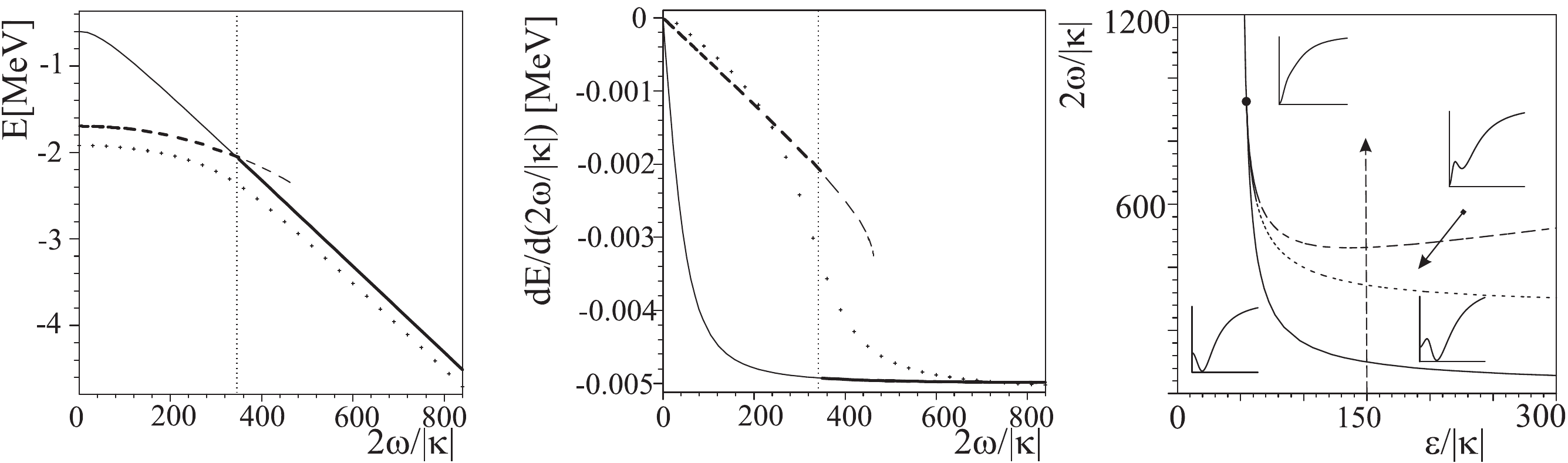}
\caption{Left panel:
Energy of the extrema of the potential~(\ref{eq:energysurface})
and of the exact ground state of the IBM-CM Hamiltonian~(\ref{eq:Hibm})
as a function of $\omega'$
for $N=10$ bosons,
$\Delta'=0$, $\epsilon'=150$, $\kappa=-0.01$~MeV and $\chi=0$.
The full line corresponds to the energy of
the spherical extremum (minimum or maximum at $\beta_0=0$)
and the dashed line to the energy
of the deformed minimum ($\beta_0\neq 0$).
The global minimum is indicated by a thick line.
The dots show the evolution of the exact ground-state energy and the thin vertical
dotted line indicates where the Maxwell line is crossed.
Middle panel: The first derivative with respect to $\omega'$
of the energies shown on the left.
The same conventions apply.
Right panel: The $(\epsilon',\omega')$ phase diagram
with the transition path followed in left and middle panels.}
\label{fig:firstorder}
\end{figure}
%\end{comment}
%
In the left panel of fig.~\ref{fig:firstorder} is shown
the energy of the extrema of the energy surface~(\ref{eq:energysurface})
and of the exact ground state of the IBM-CM Hamiltonian
(also called the local term of the binding energy \cite{fossion02})
as a function of $\omega'$
for $N=10$ bosons,
$\Delta'=0$, $\epsilon'=150$, $\kappa=-0.01$ MeV and $\chi=0$, 
which corresponds to the case of $U(5)$--$O(6)$ mixing.
The middle panel displays the first derivative
with respect to $\omega'$ of these quantities
for the same values of the other control parameters.
The right panel shows the path followed in the transition.
The local term of the binding energy 
gives the absolute energy of the $0^+$ ground state.
To obtain the total binding energy in the IBM,
one must also include terms in the Hamiltonian
which depend only on the number of bosons $N$~\cite{iachello87,fossion02}.
As these terms are constant for a given $N$,
they do not influence the phase diagram
and a study of the local term of the binding energy is sufficient.
In the left and middle panels the full (dashed) lines
correspond to the spherical (deformed) extrema.
From the crossing of the full and the dashed line
in the left panel of fig.~\ref{fig:firstorder},
it is clear that the global minimum is deformed 
until the transition path crosses the Maxwell line
where the spherical extremum becomes lowest.\\
Although the coherent-state formalism
can be regarded as a variational mean-field method,
the exact binding energies cannot be compared directly
with the energy of the global minimum of the energy surface.
This is due to the choice of the coherent state
which does not carry exact angular momentum $L=0$.
Therefore, the coherent state breaks the $O(3)$ symmetry of the Hamiltonian
which is respected in its exact diagonalisation~\cite{ginocchio80}.
Hence, rather than an exact, quantitative comparison of the energies,
the purpose of fig.~\ref{fig:firstorder} is to reveal qualitative similarities
along the transition path.
For a quantitative comparison with the exact ground-state energies,
the energy surface~(\ref{eq:energysurface}) must be projected onto $L=0$ .
From fig.~\ref{fig:firstorder} it is clear
that the exact ground-state energy and the energy of the global minimum
evolve similarly with changing control parameter $\omega'$.
The middle panel of fig.~\ref{fig:firstorder}
illustrates that the derivative of energy of the global minimum (thick line)
exhibits a discontinuity at the Maxwell line
where the energy surface thus undergoes a first-order shape phase transition
Since the derivative of the exact ground-state energy
is reasonably close to the derivative of the global minimum
and changes rapidly in the neighbourhood of the Maxwell point,
we may associate this jump in the derivative of the binding energy
with a first-order shape phase transition.
Summarising, in the case of $U(5)$--$O(6)$ mixing,
the nuclear system undergoes a first-order quantum phase transition
when passing through the line of Maxwell points.\\
%
%\begin{comment}
\begin{figure}%[!tb]
\includegraphics [width=\textwidth]{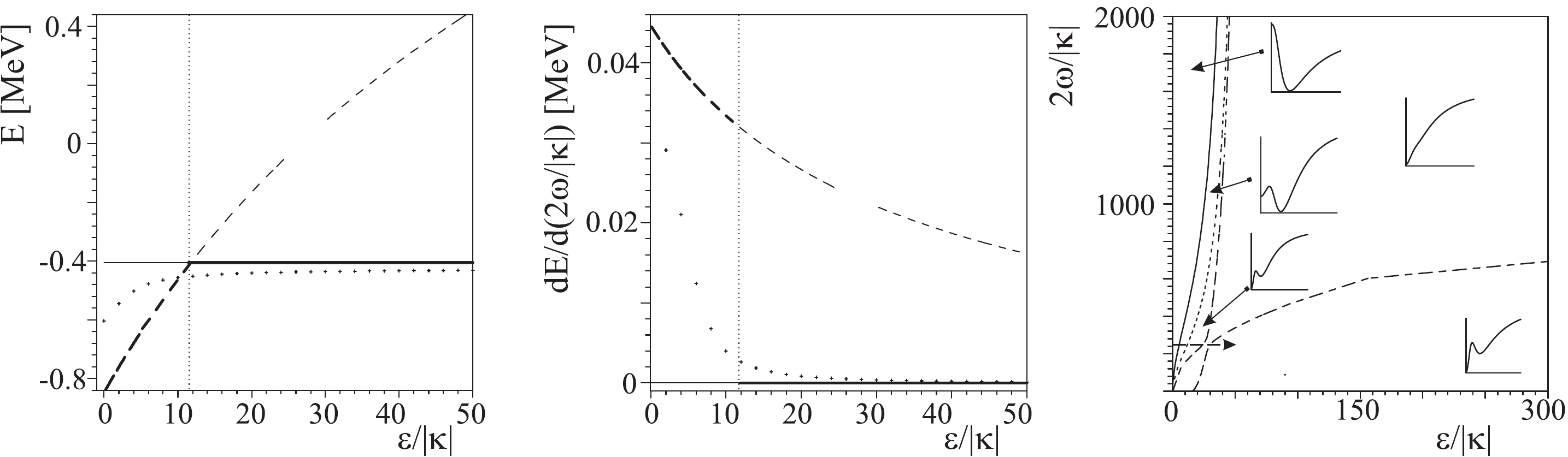}
\caption{Left panel:
Energy of the extrema of the potential~(\ref{eq:energysurface})
and of the exact ground state of the IBM-CM Hamiltonian~(\ref{eq:Hibm})
as a function of $\epsilon'$
for $N=10$ bosons,
$\Delta'=406$, $\omega'=250$, $\kappa=-0.01$~MeV
and $\chi=-\sqrt{7}/2$.
The full line corresponds to the energy of
the spherical extremum (minimum or maximum at $\beta_0=0$)
and the dashed line to the energy
of the deformed minimum ($\beta_0\neq 0$).
The global minimum is indicated by a thick line.
The dots show the evolution of the exact ground-state energy and the thin vertical
dotted line indicates where the Maxwell line is crossed.
Middle panel: The first derivative with respect to $\epsilon'$
of the energies shown on the left.
The same conventions apply.
Right panel: The $(\epsilon',\omega')$ phase diagram
with the transition path followed in left and middle panels.}
\label{fig:firstorder2}
\end{figure}
%\end{comment}
%
The same conclusion is reached for $U(5)$--$SU(3)$ mixing.
In the left panel of fig.~\ref{fig:firstorder2} is shown
the energy of the extrema of the potential surface~(\ref{eq:energysurface})
and of the exact ground state of the IBM-CM Hamiltonian
as a function of $\epsilon'$
for $N=10$ bosons,
$\Delta'=406$, $\omega'=250$, $\kappa=-0.01$~MeV
and $\chi=-\sqrt{7}/2$, 
which corresponds to the case of $U(5)$--$SU(3)$ mixing.
The middle panel displays the first derivative
with respect to $\epsilon'$ of these quantities
for the same values of the other control parameters.
The right panel shows the path followed in the transition.
In fig.~\ref{fig:firstorder2} the same conventions are followed
as in fig.~\ref{fig:firstorder};
in particular, in the left and middle panels the full (dashed) lines
correspond to the spherical (deformed) extrema.
Note that the energies and their derivatives
are studied as a function of $\epsilon'$
whereas the varying control parameter in the case of $U(5)$--$O(6)$ mixing
was $\omega'$.
The fact that the energy surface needs to be projected on $L=0$
for a quantitative comparison with the exact energies to be valid,
is immediately clear from the left panel of fig.~\ref{fig:firstorder2}.
In the deformed region for small $\epsilon'$,
the exact energy is higher than the energy of the deformed minimum.
This can be understood by realising
that the deformation-driving part of the coherent state
is also the $O(3)$-symmetry breaking part.
Hence, the need for restoring the $O(3)$ symmetry
by means of angular momentum projection
is largest in the deformed region.
Nevertheless, the global behaviour of the exact energy
and the energy of the global minimum is similar.
The discontinuity in the derivative of the energy of the latter
(see middle panel of fig.~\ref{fig:firstorder2})
again leads to the conclusion that the energy surface
undergoes a first-order phase transition
when passing through the Maxwell point.
Similarly, the slope of the derivative of the exact energy
changes strongly in the neighbourhood of the Maxwell point
and can be associated with a first-order shape phase transition
of the energy surface.
Note that the curve for the deformed minimum exhibits a gap
corresponding to the passage through the narrow region
with a single spherical minimum in the phase diagram for $\Delta'=406$.\\
The cases discussed above are specific examples
and can be repeated for any set of control parameters
passing through a region with shape coexistence. 
\subsection{Second-order phase transitions}
In the case of only one configuration,
it is known that the transition from $U(5)$ to $SU(3)$
is characterised by a first-order shape phase transition.
The transition from $U(5)$ to $O(6)$ on the other hand
is of second order~\cite{dieperink80,dieperink80bis}
As the energy surface associated with the IBM-CM
exhibits a similar transition in the case of $U(5)$--$O(6)$ mixing,
one expects this transition to be of second order.
Second-order shape phase transition
are recognised from the discontinuity in the second derivative
of the energy of the global minimum with respect to a control parameter.
Unfortunately, an analysis similar to the one for first-order shape phase transitions
experiences numerical difficulties in the neighbourhood of a critical point.
Therefore, we use a different method to identify this transition.
It is possible to recognise second-order phase transitions
by the observation of a power law behaviour of physical quantities
when passing through a critical point.
In general, a power law describes the power behaviour of a physical quantity
in the neighbourhood of the critical point,
\begin{equation}
F\sim |a_{\rm c}-a|^\delta
\end{equation}
where $F$ is an order parameter (observable) of the system,
$a$ is a control parameter,
$a_{\rm c}$ the value of the control parameter at the critical point
and $\delta$ a critical exponent.
This behaviour is a fingerprint of second-order phase transitions
and it is a remarkable fact that phase transitions
arising in different physical systems
often possess the same set of critical exponents.
This phenomenon is known as universality.\\
For the specific case of phase diagrams in configuration-mixed systems,
we study the evolution of the deformation $\beta_0$
at the global minimum of the energy surface
as a function of the control parameters~\cite{iachello04}.
We do this for $U(5)$--$O(6)$ mixing
and calculate the relevant critical exponents.
The main reason for the choice of this order parameter
is that the evolution of $\beta_0$ can be treated analytically.\\
The value of $\beta_0$ at the global minimum
results from solving $\partial E_-/\partial\beta=0$
in the unknowns ($\beta,\Delta',\epsilon',\omega',N$).
This equation gives rise to the following relation
between the control and order parameters:
\begin{eqnarray}
\omega'_\pm&=&
\pm\frac{4\sqrt{-(1+\beta^2)\epsilon' N(N+2)[(N+2)\beta^2-N]}}
{(1+\beta^2)^2\big([\epsilon' N+4(N+2)^2]\beta^2+
N[\epsilon' -4(N+2)]\big)}
\nonumber\\
&&\times\left([\epsilon' N-4(N+2)+\zeta]\beta^4+
[\epsilon'N+4N(N+2)+2\zeta]\beta^2+\zeta\right)~.
\label{eq:omega}
\end{eqnarray}
This relation is derived in Appendix \ref{ap:omega}.
Since the critical points separating the spherical and deformed phases
are situated on the locus of analytically obtained critical points~(\ref{eq:2fold}), we focus our attention on the point $\beta=0$ and $\epsilon'=\epsilon'_{\rm c}$
(see eq.~(\ref{eq:2fold})).
For $\zeta\neq0$ (and thus $\epsilon'_{\rm c}\neq4(N+2)$)
$\omega'_\pm$ is continuous in the point $\beta=0$
and $\epsilon'=\epsilon'_{\rm c}$.
Hence, the sign of $\omega'_\pm$ in ($\beta=0$, $\epsilon'=\epsilon'_{\rm c}$)
remains unchanged in a sufficiently small region around ($\beta=0$, $\epsilon'=\epsilon'_{\rm c}$).
In this point $\omega'_\pm$ reduces to
\begin{equation}
\omega'_\pm(\beta=0,\epsilon'=\epsilon'_{\rm c})=
\pm\frac{4\zeta\sqrt{\epsilon'_{\rm c}(N+2)}}
{\epsilon'_{\rm c}-4(N+2)}~,
\end{equation}
Because $\epsilon'=4(N+2)$ is the vertical asymptote
in the case of $U(5)$--$O(6)$ mixing,
$\epsilon'_{\rm c}<4(N+2)$ if $\zeta<0$
and $\epsilon'_{\rm c}>4(N+2)$ if $\zeta>0$ (See sect. \ref{sec:analytical}).
Hence, in the neighbourhood of ($\beta=0$, $\epsilon'=\epsilon'_{\rm c}$)
$\omega'_{+}$ is positive. Expanding $\omega'_{+}$ 
around ($\beta=0$, $\epsilon'=\epsilon'_{\rm c}$),
we find to lowest order
\begin{eqnarray}
\omega'_+&=&
\omega'_+(\beta=0,\epsilon'=\epsilon'_{\rm c})+\frac{\partial\omega'_+}{\partial\beta}\bigg|_{\beta=0,\epsilon'=\epsilon'_{\rm c}}\beta+\frac{\partial\omega'_+}{\partial\epsilon'}\bigg|_{\beta=0,\epsilon'=\epsilon'_{\rm c}}(\epsilon'-\epsilon'_{\rm c})\nonumber\\
&&+\frac{1}{2}\frac{\partial^2\omega'_+}{\partial\beta^2}\bigg|_{\beta=0,\epsilon'=\epsilon'_{\rm c}}\beta^2+\frac{1}{2}\frac{\partial^2\omega'_+}{\partial\epsilon'^2}\bigg|_{\beta=0,\epsilon'=\epsilon'_{\rm c}}(\epsilon'-\epsilon'_{\rm c})^2 + \frac{\partial^2\omega'_+}{\partial\beta\partial\epsilon'}\bigg|_{\beta=0,\epsilon'=\epsilon'_{\rm c}}(\epsilon'-\epsilon'_{\rm c})\beta \nonumber\\
&=&
\frac{4\zeta\sqrt{\epsilon'_{\rm c}(N+2)}}{\epsilon'_{\rm c}-4(N+2)}
-\frac{2\zeta(N+2)[4(N+2)+\epsilon'_{\rm c}]}
{\sqrt{\epsilon'_{\rm c}(N+2)}[4(N+2)-\epsilon'_{\rm c}]^2}
(\epsilon'-\epsilon'_{\rm c})
\label{eq:taylor_omega}\\
&&-\frac{4\epsilon'_{\rm c}(N+2)[4(N+2)+\epsilon'_{\rm c}]
[4N^2(N+2)+\zeta(N+1)-\epsilon'_{\rm c} N^2]}
{ N\sqrt{\epsilon'_{\rm c}(N+2)}[4(N+2)-\epsilon'_{\rm c}]^2}\beta^2~.
\nonumber
\end{eqnarray}
If we invert relation (\ref{eq:2fold})
such that $\omega'_{\rm c}$ becomes a function of $\epsilon'_{\rm c}$,
we obtain 
\begin{equation}
\omega'_{\rm c}=
\pm\frac{4\zeta\sqrt{\epsilon'_{\rm c}(N+2)}}{\epsilon'_{\rm c}-4(N+2)}~.
\end{equation}
Again, $\omega'_{\rm c}$ with the overall plus-sign is positive.
If we subsitute the positive $\omega'_{\rm c}$
in expression~(\ref{eq:taylor_omega}),
it is clear that $\omega'_{\rm c}$
and the term $\omega'_+(\beta=0,\epsilon'=\epsilon'_{\rm c})$ cancel
and expression~(\ref{eq:taylor_omega}) reduces to
\begin{multline}
\frac{4\epsilon'_{\rm c}(N+2)[4(N+2)+\epsilon'_{\rm c}]
[4N^2(N+2)+\zeta(N+1)-\epsilon'_{\rm c} N^2]}
{N\sqrt{\epsilon'_{\rm c}(N+2)}[4(N+2)-\epsilon'_{\rm c}]^2}\beta^2
=\\
\frac{2\zeta(N+2)[4(N+2)+\epsilon'_{\rm c}]}
{\sqrt{\epsilon'_{\rm c}(N+2)}[4(N+2)-\epsilon'_{\rm c}]^2}
(\epsilon'_{\rm c}-\epsilon')~.
\end{multline}
We conclude that the deformed minimum occurs with deformation 
\begin{multline}
\beta_0=
\sqrt{\frac{\zeta N}
{2\epsilon'_{\rm c}\left[4N^2(N+2)+\zeta (N+1)-\epsilon'_{\rm c} N^2\right]}}
(\epsilon'_{\rm c}-\epsilon')^{1/2}
\\
\textmd{for $\omega'_{\rm c}>\omega'_{\rm t}$
and $\epsilon'<\epsilon'_{\rm c}$}~,
\label{eq:criticalexponent1}
\end{multline}
in the neighbourhood of the critical point
($\epsilon'_{\rm c},\omega'_{\rm c},\Delta',N$).
The conditions $\omega'_{\rm c}>\omega'_{\rm t}$
and $\epsilon'<\epsilon'_{\rm c}$ ensure
that the deformation $\beta_0$ is real.
It is clear that in the neighbourhood of the critical point
($\epsilon'_{\rm c},\omega'_{\rm c},\Delta',N$),
the deformation $\beta_0$ at the deformed minimum
exhibits a power law behaviour with a critical exponent 1/2.\\
If $\zeta=0$ then $\epsilon'_{\rm c}$ takes the constant value $4(N+2)$,
such that $\omega'_\pm$ in~(\ref{eq:omega})
is discontinuous in ($\beta=0$, $\epsilon'=\epsilon'_{\rm c}$).
From appendix \ref{ap:omega}, it follows that $\omega'$ is undefined 
in ($\beta=0$, $\epsilon'=\epsilon'_{\rm c}$), hence, it is not possible 
to derive an analytical expression for the powerlaw.\\
%
%\begin{comment}
\begin{figure}%[!t]
\begin{center}
\includegraphics[width=0.7\textwidth]{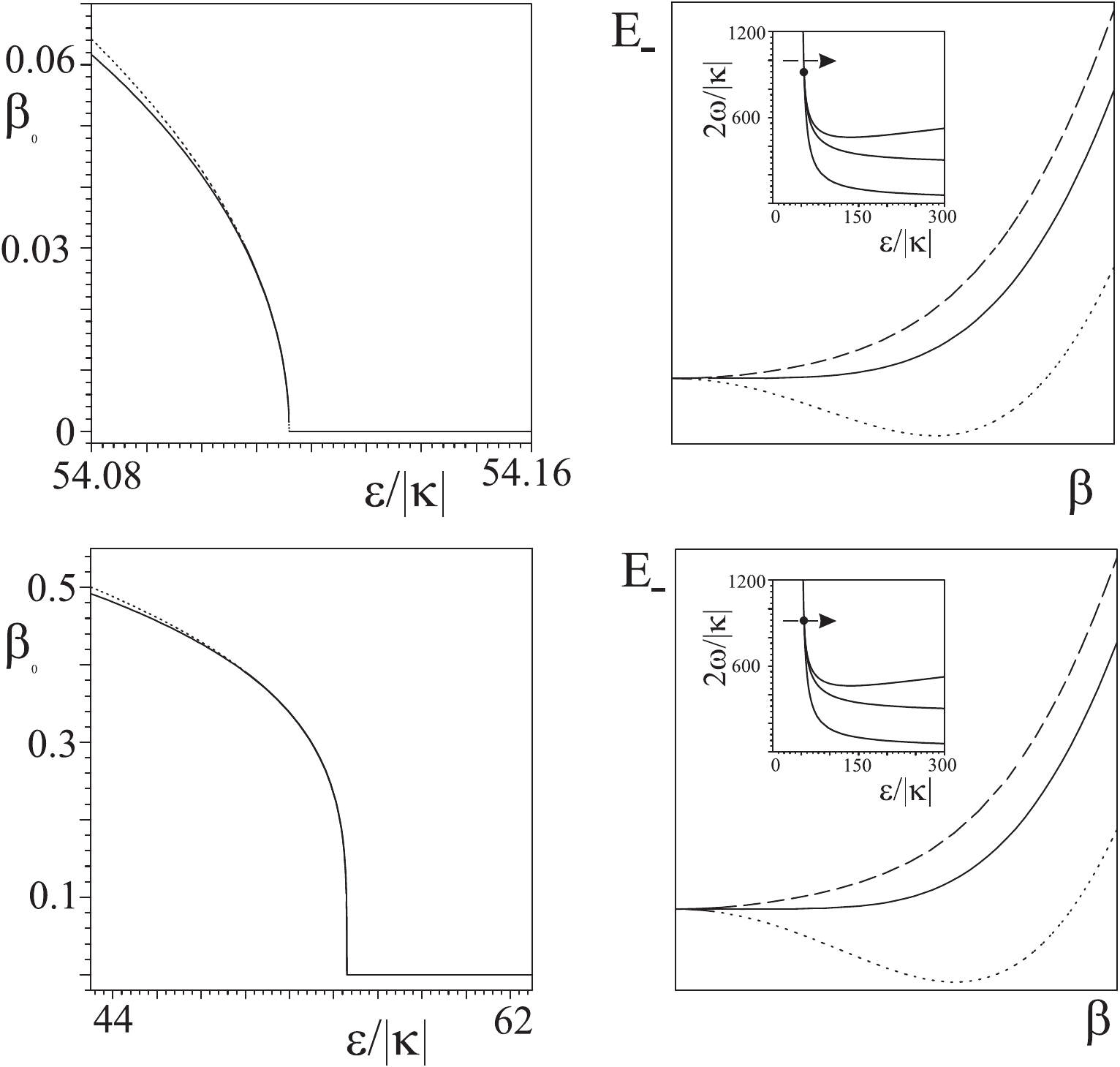}
\end{center}
\caption{
Power law behaviour at the critical point (upper panels)
and at the triple point (lower panels).
The left panels shows the deformation at minimum, $\beta_0$,
as a function of $\epsilon'$ as predicted by the power law (dotted line)
and as compared with a numerical calculation
(full line).
The right panels show the energy surface as a function of $\beta$,
where the dotted line corresponds
to $\epsilon'<\epsilon'_{\rm c}$ or $\epsilon'_{\rm t}$,
the full line to $\epsilon'$ at the critical or triple point,
and the dashed line to $\epsilon'>\epsilon'_{\rm c}$ or $\epsilon'_{\rm t}$.
The inset plots indicate the transition path in the phase diagram.
Other control parameters are $N=10$, $\Delta'=0$,
and $\omega'=1000$ (upper panels)
or $\omega'=\omega'_{\rm t}\approx930.7954$ (lower panels).
}
\label{fig:powerlaw}
\end{figure}
%\end{comment}
%
In the upper panels of fig.~\ref{fig:powerlaw}
the deformation at minimum, $\beta_0$, is shown
for $N=10$ bosons, $\omega'=1000$ and $\Delta'=0$
as $\epsilon'$ crosses the critical line of second order.
The evolution of $\beta_0$ in the neighbourhood of the critical point
is compared with a power law with critical exponent 1/2
as obtained in~(\ref{eq:criticalexponent1}) in the left panel.
The energy surfaces just before and just after crossing the critical point
as well as at the critical point are shown in the right upper panel
The inset figure shows the path followed in the phase diagram.
The comparison of the power law~(\ref{eq:criticalexponent1})
and the exact $\beta_0$ demonstrates the validity of the former
in the neighbourhood of the critical point.\\
In the triple point itself (see eq.~(\ref{eq:triplepoint}))
the coefficient of $\beta^2$ in the Taylor expansion
of $\omega'_+$ in~(\ref{eq:omega}) vanishes
and higher-order terms have to be considered.
If $\zeta\neq0$, the Taylor expansion
around $\beta=0,\epsilon'=\epsilon'_{\rm t}$
becomes to lowest order in $\beta$ and $\epsilon'-\epsilon'_{\rm t}$
\begin{equation}
\omega'_+=
a_{00}+
a_{01}(\epsilon'-\epsilon'_{\rm t})+
a_{02}(\epsilon'-\epsilon'_{\rm t})^2+
a_{21}(\epsilon'-\epsilon'_{\rm t})\beta^2+
a_{03}(\epsilon'-\epsilon'_{\rm t})^3+
a_{40}\beta^4,
\label{eq:taylor2}
\end{equation} 
where
\begin{align}
&a_{00}=\frac{4N}{N+1}\sqrt{(N+2)[4N^2(N+2)+\zeta(N+1)]}~,
\nonumber\\
&a_{01}=-\frac{2N^3(N+2)[8N^2(N+2)+\zeta(N+1)]}
{\zeta(N+1)^2\sqrt{(N+2)[4N^2(N+2)+\zeta(N+1)]}}~,
\nonumber\\
&a_{02}=\frac{N^5(N+2)^2
\big(3[8N^2(N+2)+\zeta(N+1)]^2-64N^4(N+2)^2\big)}
{2\zeta^2(N+1)^3\big((N+2)[4N^2(N+2)+\zeta(N+1)]\big)^{3/2}}~,
\nonumber\\
&a_{21}=\frac{4(N+2)^2N^2[8N^2(N+2)+\zeta(N+1)][4N^2(N+2)+\zeta(N+1)]^2}{\zeta^2(N+1)^2\big((N+2)[4N^2(N+2)+\zeta(N+1)]\big)^{3/2}}~,
\nonumber\\
&a_{03}=-\frac{N^7(N+2)^3}{4\zeta^3(N+1)^4
\big((N+2)[4N^2(N+2)+\zeta(N+1)]\big)^{5/2}}
\nonumber\\
&\qquad\times\left(5[8N^2(N+2)+\zeta(N+1)]^3-64N^4(N+2)^2[5\zeta(N+1)+24N^2(N+2)]\right)~,
\nonumber\\
&a_{40}=-\frac{6(N+2)^2[8N^2(N+2)+\zeta(N+1)][4N^2(N+2)+\zeta(N+1)]^2}
{N\zeta\big((N+2)[4N^2(N+2)+\zeta(N+1)]\big)^{3/2}}~.
\end{align}
For $\omega'=\omega'_{\rm t}$ the left-hand side of eq.~(\ref{eq:taylor2})
cancels with $a_{00}$
and the resulting equation
can be solved as a quadratic equation in $\beta^2$.
Keeping only the leading term in $(\epsilon'-\epsilon'_{\rm t})$,
we derive the following expression
for the deformation $\beta_0$ at the minimum
in the neighbourhood of the triple point:
\begin{eqnarray}
\beta_0&=&
\left(\frac{a_{01}}{a_{40}}\right)^{1/4}(\epsilon'_{\rm t}-\epsilon')^{1/4}~,
\\&=&
\left(\frac{N^4}{3(N+1)^2[4N^2(N+2)+\zeta(N+1)]}\right)^{1/4} (\epsilon'_{\rm t}-\epsilon')^{1/4}~.
\label{eq:criticalexponent2}
\end{eqnarray}
Thus, at the triple point the critical exponent changes from 1/2 to 1/4.
The behaviour of $\beta_0$ at the triple point
and its comparison with the power law of eq.~(\ref{eq:criticalexponent2})
is shown in the lower panels of fig.~\ref{fig:powerlaw}.
Again, the comparison is very good.
Note that the critical exponents for the order parameter $\beta_0$
are the same as those in Landau theory of tricritical points~\cite{plischke94}.
\section{Conclusion}
In the past years, many theoretical and experimental studies
have focused on the subject of quantum phase transitions in atomic nuclei.
The interacting boson model provides a tractable framework
to study quantum phase transitions from a theoretical point of view.
Because of the algebraic foundations of the model,
an energy surface is easily constructed
and can be studied within the framework of catastrophe theory
allowing its qualitative study as a function of the control parameters.\\
In the present work a detailed study of the energy surface
associated with configuration mixing between a spherical $U(5)$
and a deformed $\hat{Q}(\chi)\cdot\hat{Q}(\chi)$ configuration was performed.
By expanding the energy surface around $(\beta,\gamma)=(0,n\pi/3)$
we have derived an analytical solution of the criticality conditions.
An analytical expression for the triple point was obtained
and it was shown that it only occurs in the case of $U(5)$--$O(6)$ mixing.
For general $\beta\neq0$ the criticality conditions
must be solved for numerically.
The same holds for the Maxwell points
which indicate where the global minimum
jumps from one deformation to another.
Phase diagrams for the two most symmetrical cases
of $U(5)$--$O(6)$ and $U(5)$--$SU(3)$ mixing
have been constructed an discussed.
Both cases display a large region of shape coexistence
for a broad range of excitation energies of the intruder configuration.
For very high excitation energies
the presence of the triple point in the case of $U(5)$--$O(6)$
implies the disappearance of the region of shape coexistence
for low $\epsilon/|\kappa|$
whereas this region is always present
for other deformed intruder configurations ({\it i.e.}, for $\chi\neq0$).\\
Finally, we have discussed the order of the shape phase transitions.
It turns out that, generally, the derivative of the energy
of the global minimum of the energy surface
changes discontinuously at the Maxwell line
and undergoes a first-order shape phase transition.
In a numerical calculation the derivative of the binding energy
follows this behaviour
although the discontinuity is smoothed out because of finite-size effects.
For the transition from a spherical to a deformed minimum
in the case of $U(5)$--$O(6)$ mixing,
we have shown that the deformation $\beta_0$ of the global minimum
exhibits a power-law behaviour in the neighbourhood of the critical point
and we have given analytical expressions for the critical exponents.
Hence this transition is of second order.
\section*{Acknowledgements}
The authors are grateful to A.~Frank, P.~Cejnar and J.~Ryckebusch
for interesting discussions.
Financial support from the ``FWO-Vlaanderen" (V.H and K.H.)
and the University of Ghent (S.D.B. and K.H.)
which made this research possible, is acknowledged. V.H. and S.D.B. also
received financial support from the European Union under contract No 2000-00084. K.H. 
likes to thank the ISOLDE group for their hospitality during the final 
stage of this work.
\begin{appendix}
\section{Analytical solution of $\partial E_-/\partial\beta=0$
in the case of $U(5)$--$O(6)$ mixing}
\label{ap:omega}
In this appendix we derive the expression~(\ref{eq:omega})
which results from solving the condition $\partial E_-/\partial\beta=0$.
Substituting $\chi=0$, $\Delta'=-\zeta+5(N+2)$,
$\omega'=2\omega/|\kappa|$ and
\begin{eqnarray}
a_1&=&\epsilon'N+4(N+2)-\zeta~,
\nonumber\\
a_2&=&\epsilon'N-4N(N+2)-2\zeta~,
\nonumber\\
b_1&=&\epsilon'N-4(N+2)+\zeta~,
\nonumber\\
b_2&=&\epsilon'N+4N(N+2)+2\zeta~,
\end{eqnarray} 
in the expression for the energy surface~(\ref{eq:energysurface}), we find
\begin{equation}
E_-=
\frac{|\kappa|}{2(1+\beta^2)^2}
\left(a_1\beta^4+a_2\beta^2-\zeta-
\left[\left(b_1\beta^4+b_2\beta^2+\zeta\right)^2
+\omega'^2(1+\beta^2)^4\right]^\frac{1}{2}\right)~.
\end{equation}
Upon a scaling factor $|\kappa|/2$
the first derivative $\partial E_-/\partial\beta$
can be written as
\begin{eqnarray}
\frac{\partial E_-}{\partial\beta}&=&-\frac{4\beta}{(1+\beta^2)^3}\left(a_1\beta^4+a_2\beta^2-\zeta-\left[(b_1\beta^4+b_2\beta^2\right.\right.\nonumber\\
&+&\left.\left.\zeta)^2+\omega'^2(1+\beta^2)^4\right]^{\frac{1}{2}}\right)+\frac{4a_1\beta^3+2a_2\beta}{(1+\beta^2)^2}\nonumber\\
&-&\frac{(b_1\beta^4+b_2\beta^2+\zeta)(4b_1\beta^3+2b_2\beta)+4\omega'^2(1+\beta^2)^3\beta}{(1+\beta^2)^2\left[(b_1\beta^4+b_2\beta^2+\zeta)^2+\omega'^2(1+\beta^2)^4\right]^{\frac{1}{2}}}~.
\end{eqnarray}
This expression can be rewritten to
\begin{eqnarray}
\frac{\partial E_-}{\partial\beta}
&=&\frac{2\beta}
{(1+\beta^2)^3\left[(b_1\beta^4+b_2\beta^2+\zeta)^2+
\omega'^2(1+\beta^2)^4\right]^{\frac{1}{2}}}
\nonumber\\
&\times&
\bigg(\left[(2a_1-a_2)\beta^2+(a_2+2\zeta)\right]
\left[(b_1\beta^4+b_2\beta^2+\zeta)^2+
\omega'^2(1+\beta^2)^4\right]^{\frac{1}{2}}
\nonumber\\
&&+(b_1\beta^4+b_2\beta^2+\zeta)
\left[(b_2-2b_1)\beta^2+(-b_2+2\zeta)\right]\bigg)~.\label{eq:derivative}
\end{eqnarray}
Assuming that $(2a_1-a_2)\beta^2+(a_2+2\zeta)\neq 0$
and $\beta\neq 0$, the condition $\partial E_-/\partial\beta=0$ leads to
\begin{eqnarray}
\omega'^2(1+\beta^2)^4&=&
\frac{(b_1\beta^4+b_2\beta^2+\zeta)^2
\left[(2b_1-b_2)\beta^2+(b_2-2\zeta)\right]^2}
{\left[(2a_1-a_2)\beta^2+(a_2+2\zeta)\right]^2}
\nonumber\\
&&-(b_1\beta^4+b_2\beta^2+\zeta)^2~,
\end{eqnarray}
This can be rewritten as
\begin{eqnarray}
\omega'^2&=&
\frac{(b_1\beta^4+b_2\beta^2+\zeta)^2}
{(1+\beta^2)^4\left[(2a_1-a_2)\beta^2+(a_2+2\zeta)\right]^2}
\big([2(b_1-a_1)-(b_2-a_2)]\beta^2
\nonumber\\
&&+(b_2-a_2-4\zeta)\big)
\big([2(b_1+a_1)-(b_2+a_2)]\beta^2+(b_2+a_2)\big)~.
\end{eqnarray}
Inserting the expressions for $a_1, a_2, b_1$ and $b_2$,
we find eq.~(\ref{eq:omega}):
\begin{eqnarray}
\omega'_\pm&=&
\pm\frac{4\sqrt{-(1+\beta^2)\epsilon' N(N+2)[(N+2)\beta^2-N]}}
{(1+\beta^2)^2\big([\epsilon' N+4(N+2)^2]\beta^2+
N[\epsilon' -4(N+2)]\big)}
\nonumber\\
&&\times\left([\epsilon' N-4(N+2)+\zeta]\beta^4+
[\epsilon'N+4N(N+2)+2\zeta]\beta^2+\zeta\right)~.
\end{eqnarray}
If $\beta=0$, the condition $\partial E_-/\partial\beta=0$ is automatically fullfilled. Hence, there is always an extremum,
either a minimum or a maximum, at $\beta_0=0$.
Finally, if $(2a_1-a_2)\beta^2+(a_2+2\zeta)=0$, it follows that
\begin{equation}
\epsilon'_i=-\frac{4(N+2)[(N+2)\beta_i^2-N]}{N(1+\beta_i^2)}~.
\label{eq:epsspec}
\end{equation}
Inserting this relation in $\partial E_-/\partial\beta=0$,
we derive the following expression for $\zeta$ from the condition
$(b_1\beta^4+b_2\beta^2+\zeta)=0$ (see eq. \ref{eq:derivative}) 
\begin{equation}
\zeta_i=\frac{4(N+2)\beta_i^2[(N+3)\beta_i^2-2N]}{(1+\beta_i^2)^2}~.
\label{eq:zetaspec}
\end{equation}
The index $i$ has been added to demonstrate
that $\epsilon_i$, $\beta_i$ and $\zeta_i$ cannot be varied independently.
The parameter $\omega'$ however can take on any value. Hence, for a given $\zeta_i$, 
there is a corresponding $\epsilon_i$ for which the deformation $\beta_i$ of the extremum 
of the energy surface remains unchanged when $\omega'$ is varied.\\
Another solution to $\partial E_-/\partial\beta=0$ follows from the condition
 $(b_2-2b_1)\beta^2+(-b_2+2\zeta)=0$ (see eq. \ref{eq:derivative}). The deformation
$\beta_i$ then equals $\pm\sqrt{N/(N+2)}$ and $\epsilon_i$ (\ref{eq:epsspec}) becomes zero.

\end{appendix}

\bibliographystyle{h-elsevier}

\end{document}